\documentclass[11pt,draftcls,onecolumn]{IEEEtran}
\usepackage{amsfonts}
\usepackage{amssymb}
\usepackage{mathrsfs}
\usepackage{amsmath}
\usepackage{cite}
\usepackage{mathrsfs}
\usepackage[ruled,vlined]{algorithm2e}
\usepackage{tikz}
\usetikzlibrary{arrows,automata}
\usepackage[latin1]{inputenc}
\usepackage{verbatim}
\makeatletter

\newcommand{\Rmnum}[1]{\expandafter\@slowromancap\romannumeral #1@}
\makeatother

\newtheorem{thm}{Theorem}
\newtheorem{lemma}[thm]{Lemma}

\newtheorem{prop}{Proposition}

\newtheorem{defn}{Definition}
\newtheorem{rem}[thm]{Remark}

\newcommand{\w}{{\omega}}
\newcommand{\vb}{\vec{b}}
\newcommand{\vf}{\vec{f}}
\newcommand{\vk}{\vec{k}}

\newcommand{\vm}{\vec{m}}
\newcommand{\vc}{\vec{c}}
\newcommand{\Fq}{\mathbb{F}_q}

\newcommand{\mA}{\mathcal{A}}
\newcommand{\mC}{\mathcal{C}}

\newcommand{\mN}{\mathcal{N}}
\newcommand{\bzero}{{\bf 0}}
\newcommand{\wtE}{\widetilde{E}}
\newcommand{\Rank}{{\mathrm{Rank}}}
\newcommand{\tail}{{\mathrm{tail}}}
\newcommand{\head}{{\mathrm{head}}}
\newcommand{\Out}{{\mathrm{Out}}}
\newcommand{\In}{{\mathrm{In}}}
\newcommand{\cut}{{\mathrm{cut}}}
\newcommand{\mincut}{{\mathrm{mincut}}}

\newcommand{\mfK}{\mathfrak{K}}

\begin{document}

\title{On the Optimality of Secure Network Coding
\thanks{This research is supported by the National Key Basic Research Program of China (973 Program Grant No. 2013CB834204), the National Natural Science
Foundation of China (Nos. 61301137, 61171082) and the Fundamental Research Funds for Central Universities of China (No. 65121007).}}

\author{Xuan~Guang~\IEEEmembership{Member,~IEEE},
        Jiyong~Lu~\IEEEmembership{Student Member,~IEEE},
        and~Fang-Wei~Fu~\IEEEmembership{Member,~IEEE}
\thanks{X. Guang is with the School of Mathematical Sciences and LPMC, J. Lu and F.-W. Fu are with the Chern Institute of
Mathematics, Nankai University, Tianjin 300071, China (e-mail:
xguang@nankai.edu.cn, lujiyong@mail.nankai.edu.cn, fwfu@nankai.edu.cn).}}

\markboth{}%
{}
%


\maketitle

\begin{abstract}
In network communications, information transmission often encounters wiretapping attacks. Secure network coding is introduced to prevent information from being leaked to adversaries. The investigation of performance bounds on the numbers of source symbols and random symbols are two fundamental research problems. For an important case that each wiretap-set with cardinality not larger than $r$, Cai and Yeung proposed a coding scheme, which is optimal in the senses of maximizing the number of source symbols and at the same time minimizing the number of random symbols. In this letter, we further study achievable lower bound on the number of random key and show that it just depends on the security constraint, and particularly, is independent to the information amount for transmission. This implies that when the number of transmitted source message changes, we can't reduce the number of random key to keep the same security level. We further give an intuitive interpretation on our result. In addition, a similar construction of secure linear network codes is proposed, which achieves this lower bound on the number of random key no matter how much information is transmitted. At last, we also extend our result to imperfect security case.
\end{abstract}
\begin{IEEEkeywords}
Secure network coding, perfect security, imperfect security, rate of the key, achievability
\end{IEEEkeywords}

%
\IEEEpeerreviewmaketitle

\section{Introduction}

\IEEEPARstart{N}{etwork} coding allows internal nodes in a communication network to process the received information \cite{Ahlswede-Cai-Li-Yeung-2000}, which can maximize multicast throughput of the communication network. Li \textit{et al.} \cite{Li-Yeung-Cai-2003} further indicated that linear network coding is sufficient for multicast. Jaggi \textit{et al.} \cite{co-construction} proposed a deterministic polynomial-time algorithm for constructing a linear network code (LNC).

In practical network communications, information transmission is often under wiretapping attacks. An eavesdropper is capable of wiretapping on an unknown set of channels in networks. The secure network coding was introduced first by Cai and Yeung \cite{secure-conference} to prevent information from being leaked to the eavesdropper. In their journal paper \cite{Cai-Yeung-SNC-IT}, they proposed the model of a communication system on a wiretap network (CSWN) and the idea of secure network coding, where information-theoretic security \cite{Cai-Yeung-SNC-IT} (also refer to perfect security) is under consideration, i.e., the mutual information between source message and the message available to the eavesdropper is zero. Particularly, a network code is $r$-secure if this eavesdropper can obtain nothing by accessing any $r$ channels. This number $r$ is called \textit{security-level} in this letter. In addition, if it is allowed that the eavesdropper can obtain a controlled amount of information about source message, this is called imperfect security. They also proposed a coding scheme to construct a secure linear network code (SLNC) with security-level $r$, and indicated the optimality, that is, the constructed SLNCs multicast the maximum possible number of information symbols to sink nodes securely and at the same time use the minimum number of random symbols to achieve security-level $r$ when the source message is distributed uniformly. Usually, these random symbols are regarded as the random key.
Subsequently, Feldman \textit{et al.}\cite{Feldman} derived a tradeoff between the size of source message set and code alphabet size. In \cite{Rouayheb-IT}, Rouayheb \textit{et al.} shows that this secure network coding problem can be regarded as a network generalization of the wiretap channel of type \Rmnum{2} introduced by Ozarow and Wyner \cite{wiretap-channel-II}. Thus, they presented a construction of SLNCs by using secure codes for wiretap channel \Rmnum{2}, which is actually equivalent to Cai and Yeung's construction.

Similar to classical wiretap channel model in \cite{Shannon-secrecy}, in our wiretap network model, it is necessary to randomize the source message to combat the eavesdropper. Thus, randomness is introduced, which reduces the throughput inevitably. Hence, two of fundamental problems in secure network coding theory are the performance bounds on the size of source message and the size of random key under the certain security constraint. To be specific, on the one hand, we are interested in the maximum size of source message, which indicates the maximum number of information symbols allowed to be multicast to sinks from the source node securely. On the other hand, we are also interested in the minimum size of random key, which describes the minimum number of random symbols injected into networks by the source node to guarantee the required security. The former characterizes the effectiveness of SLNCs for information transmission and the latter characterizes the cost of SLNCs for the security. In general, the latter is more important and necessary since in cryptography randomness is regarded as a resource.

In \cite{Cheng-Yeung-Performance-SNC}, Cheng and Yeung investigated the above two fundamental bounds for the general case that the collection $\mA$ of all possible wiretap sets can consist of arbitrary subsets of channels, which perhaps are the most general results until now. Cai and Yeung \cite{Cai-Yeung-SNC-IT} also studied these two fundamental bounds for a special but very important case that all wiretap sets have cardinalities not larger than a fixed positive integer $r$, in other words, the eavesdropper can access any $r$ channels at most. They designed a construction of SLNCs of $\w$-rate and $r$-security-level with $\w+r=C_{\min}\triangleq \min_{t\in T}C_t$ with $C_t$ being the minimum cut capacity between the source $s$ and the sink $t$, and showed that their construction is optimal in the senses of maximizing the number of information symbols and at the same time minimizing the number of random symbols. Naturally, a more general problem is proposed: if the number of information symbols doesn't achieve the maximum, what is the minimum number of random symbols to ensure the same security requirement. In other words, when we reduce the number of source message to be transmitted, whether the number of random key injected into networks can decrease to keep the same security-level. The motivation of this problem is natural. For instance, in practical network communications, the source often multicasts messages at several different information rates, but the security requirement is unchanged because the wiretapping capability of the eavesdropper doesn't decrease for these distinct information rates \cite{Guang-locality-SLNC-ITW14}. Another example is that sometimes we need different security-levels even if the information rate is fixed since the security requirement is probably changed within a session. Therefore, no matter what, we always pay attention to the minimum number of random key. In this letter, we give a negative answer for the proposed question, that is, even if the source message cannot achieve the allowable maximum, the minimum number of random key to ensure the same security won't decrease. This further implies that the minimum number of random symbols injected into networks just depends on the security constraint, and independent to other parameters\footnote{In this letter, we always consider the nontrivial case that the sum of the information rate and the security-level does not exceed the minimum cut capacity between the source node and every sink node.}.

\section{Main Results}

Let $G=(V, E)$ be a directed acyclic single source multicast network, where $V$ and $E$ are the sets of nodes and channels, respectively, and denote by $s$ the single source node. We consider LNCs on this network over the base field $\Fq$, $q$ a prime power. A direct edge $e=(i,j)\in E$ stands for a channel from node $i$ to node $j$. Node $i$ is called the tail of $e$ and node $j$ is called the head of $e$, denoted by $\tail(e)$ and $\head(e)$, respectively. For a node $i$, define
\begin{align*}
\Out(i)&=\{e\in E:\ \tail(e)=i\},\\
\In(i)&=\{e \in E:\ \head(e)=i\}.
\end{align*}
We allow parallel channels between two nodes and thus assume reasonably that one field element can pass through a channel in one unit time. In defining an $n$-dimensional ($n\leq C_{\min}$) LNC on $G$, let $\In(s)$ consist of $n$ imaginary incoming channels terminating at the source node $s$, where assume that source message is transmitted to $s$ through them.
\begin{defn}[\textbf{Global Description of An LNC}]\
A $\Fq$-valued $n$-dimensional LNC on $G=(V,E)$ consists of an $n$-dimensional column vector $\vf_e$ for every channel $e\in E$ such that:
\begin{enumerate}
        \item $\vf_{e}$, $e\in \In(s)$, form the standard basis of $\Fq^n$;
        \item For other channels $e\in E$,
        \begin{align}\label{equ_ext_f}
        \vf_e=\sum_{d\in \In(\tail(e))}k_{d,e}\cdot\vf_d,
        \end{align}
        where $k_{d,e}\in \Fq$ is the local encoding coefficient for the adjacent channel pair $(d,e)$.
\end{enumerate}
\end{defn}

The source $s$ generates the random message $M$ according to uniform distribution on $\Fq^{\w}$, the message set, and $\w$ is the information rate. We request that the source message $M$ is multicast to every sink node $t\in T$, the set of sink nodes, while being protected from an eavesdropper who can access any subset of channels with cardinality not larger than $r$, i.e., the security-level is $r$. \footnote{It is necessary that $r$ is strictly smaller than the minimum cut capacity $C_t$ between the source node $s$ and every sink node $t\in T$, since otherwise an eavesdropper accessing all the channels in a minimum cut between the source node and a sink node would obtain all information received by this sink node, and thus, it can decode the source message successfully.} Let the key $K$ be an independent random variable according to uniform distribution on the alphabet $\mfK$. In order to guarantee security-level $r$, we are interested in the minimum number of the random key such that the message $M$ can be multicast from the source node $s$ to all sinks securely. In other words, we are interested in the minimum entropy $H(K)=\log |\mfK|$ of the random key $K$, or equivalently, the minimum value of the cardinality $|\mfK|$.

\begin{defn}
Define the rate of the key $K$ as:
\begin{align*}
r_K=\frac{H(K)}{\log q}.
\end{align*}
\end{defn}

Similarly, this rate of the key can also describe the number of random key $K$. For the minimum entropy of the key to guarantee security-level $r$, the corresponding optimal rate of the key is denoted by $r_K^*$.

\subsection{Perfect Security Case}

Now, we give the main conclusion for the scenario of perfect security. In \cite{Cai-Yeung-SNC-IT}, Cai and Yeung proposed a construction to design $r$-security-level and $\w\triangleq(C_{\min}-r)$-information-rate LNCs and showed that this construction maximizes the number of source symbols and minimizes the number of random symbols. This implies that for security-level $r$, the optimal rate satisfies $r_K^*=r$ for the case that $\w+r=C_{\min}$. Thus, we just consider the case $\w+r<C_{\min}$, or $\w<C_{\min}-r$. By partly using their arguments for converse part of the proof more delicately and technically, we obtain the result below.

\begin{thm}\label{thm_opt}
Let the required security-level be $r$. Whatever the information rates $\w$ satisfying $\w+r< C_{\min}$ are, the optimal rate $r_K^*$ of the key is $r$.
\end{thm}

\begin{IEEEproof}
First, recall that $G=(V,E)$ is a single source multicast network. On the one hand, we will prove the achievability, i.e., $r$ is sufficient for the rate of the key $r_K$ to construct a SLNC with information rate $\w$ and security-level $r$. Hence, in order to show the sufficiency of the key rate $r$, we want to apply the construction in \cite{Cai-Yeung-SNC-IT} to simplify the proof. However, since the summation of both information and key rates reaches the network capacity for their construction, we cannot use it directly, and thus, we will modify it to adjust to arbitrary allowed information rate by adding a constant vector $C$ to main information vector.

\noindent \textbf{Construction:}
\begin{enumerate}
  \item For the network $G$, construct an $n\triangleq C_{\min}$ dimensional LNC $\mC$ of global description $\{\vf_e: e\in E\}$ (e.g. Jaggi \emph{et al.}'s algorithm \cite{co-construction});
  \item Choose $n$ linearly independent $\Fq$-valued $n$-column vectors $\vb_1, \vb_2, \cdots, \vb_n$ satisfying the condition:
  \begin{align}\label{secure_condition}
  \langle \{ \vb_i:\ 1\leq i \leq n-r \} \rangle \cap \langle \{ \vf_e:\ e\in A \} \rangle=\{\bzero\}
  \end{align}
  for all channel-sets $A\in \wtE_r \triangleq \{ A\subseteq E:\ |A|=\Rank(F_A)=r \}$ with $F_A=\big[ \vf_e: e\in A \big]$. Then define an $n\times n$ invertible matrix
  $Q=\begin{bmatrix} \vb_1 & \vb_2 & \cdots & \vb_n \end{bmatrix}$;
  \item Let the information rate and the security-level be $\w$ and $r$, respectively, which satisfy $\w+r<n$. The source message $M$ is randomly chosen from $\Fq^\w$ according to uniform distribution, while the independent random key $K$ is distributed uniformly on $\Fq^r$.  Let $X=[M\ C \ K]$ be the input of the network $G$, where $C$ is a constant $(n-\w-r)$-dimensional row vector $\vc$. For convenience of analysis, we also regard $C$ as a random variable with probability $Pr(C=\vc)=1$.
      Furthermore, let the outcome $\vm$ of $M$ be an $\w$-row vector in $\Fq^{\w}$, and the outcome $\vk$ of $K$ be an $r$-row vector in $\Fq^r$.
  \item Encode the vector $X$ by transmitting in each channel $e$ the value $XQ^{-1}\vf_e$.
\end{enumerate}

In \cite{Cai-Yeung-SNC-IT}, the authors have shown that $\{Q^{-1}\vf_e: e\in E\}$ constitutes a global description of an $n$-dimensional SLNC, which can decode every input $[\vm\ \vc\ \vk]$ successfully and satisfy the security requirement $H(M,C|Y_A)=H(M,C)=H(M)$, further implying $H(M|Y_A)=H(M,C|Y_A)=H(M)$, for any channel-subset $A$ with cardinality not larger than $r$. Thus, we show that the optimal key rate $r_K^*$ doesn't exceed $r$, i.e., $r_K^*\leq r$.

On the other hand, we will prove that the optimal key rate $r_K^*$ is lower bounded by $r$, i.e.,  $r_K^*\geq r$, or equivalently, $H(K)\geq r\log q$.
Let $\bar{r}_K^*=\lceil r_K^* \rceil$, and clearly, $\bar{r}_K^* \geq r_K^*\geq 0$. Then there exists an injection $\phi$ from the set $\mfK$ of the key to the vector space $\Fq^{\bar{r}_K^*}$. Further, according to this injection mapping $\phi$, any output $k$ in $\mfK$ is transformed to a $\Fq$-valued $\bar{r}_K^*$-dimensional row vector $\vk$, that is, $\phi(k)=\vk$. Further, we have
$$ \w+r_K^* \leq \w+\bar{r}_K^* \leq \w+r < C_{\min}.$$
Let $CUT$ be an arbitrary channel-cut between the source node $s$ and the sink node $t$, and clearly,
$$|CUT|\geq C_{\min} > \w+\bar{r}_K^*.$$
Since the considered network code is linear, there must exist $|CUT|$ $(\w+\bar{r}_K^*)$-column vectors $\vf_e$, $e\in CUT$, such that
\begin{align*}
[M\ K] \cdot F_{CUT}=Y_{CUT},
\end{align*}
where $F_B=[\vf_e: e\in B]$ and $Y_B=[Y_e: e\in B]$ \footnote{Notice that $\vf_e$ and $Y_e$ may not be the global encoding kernel of the channel $e$ and the observation transmitted on the channel $e$, e.g., in the proof of the achievability. But they are determined exactly provided that the corresponding global encoding kernel and the observation on the channel are obtained. Thus, we still use the same symbols to represent them, and this abuse of notation should cause no ambiguity.} for any channel-set $B\subseteq E$.

Furthermore, since $\Rank(F_{CUT})\leq \w+\bar{r}_K^*$, there exists at least one channel-set $U\subseteq CUT$ with $|U|=\w+\bar{r}_K^*$ such that
$$\Rank(F_U)=\Rank(F_{CUT}),$$
and one has
\begin{align}\label{equ_1}
[M\ K] \cdot F_U=Y_U.
\end{align}
In addition, at the sink node $t\in T$, one has the decoding equation:
\begin{align*}
[M\ K] \cdot F_{\In(t)}=Y_{\In(t)},
\end{align*}
and the relationship
\begin{align*}
\langle F_{\In(t)} \rangle \subseteq \langle F_{CUT} \rangle = \langle F_{U} \rangle,
\end{align*}
where $\langle F_{B} \rangle =\langle \{ \vf_e: e\in B \} \rangle$ for any channel-set $B\subseteq E$. Hence, there exists a $|U|\times |\In(t)|=(\w+\bar{r}_K^*)\times |\In(t)|$ matrix $P$ such that
$$F_U\cdot P=F_{\In(t)}.$$
Consequently, multiplying both sides of the equality (\ref{equ_1}) by $P$ yields
\begin{align*}
[M\ K] \cdot F_{U}\cdot P=[M\ K] \cdot F_{\In(t)}=Y_{\In(t)}=Y_{U}\cdot P,
\end{align*}
and the last equality implies that
\begin{align}\label{equ_2}
H(Y_{\In(t)}|Y_U)=0.
\end{align}
Note that the sink node $t$ can recover the message $M$ with no error, which further leads to $H(M|Y_{\In(t)})=0$.
Together with the equality (\ref{equ_2}), it follows that $H(M|Y_U)=0$. Thus, for any $J\subset U$ with $|J|=r$, we obtain
\begin{align}
H(M)&=H(M|Y_U)+I(M ; Y_U)\nonumber\\
    &=I(M ; Y_U)\label{equ_3}\\
    &=I(M ; Y_{U\backslash J}, Y_J)\nonumber\\
    &=I(Y_J ; M)+I(Y_{U\backslash J}; M|Y_J)\nonumber\\
    &=I(Y_{U\backslash J}; M|Y_J)\label{equ_4},
\end{align}
where the equality (\ref{equ_3}) follows from $H(M|Y_U)=0$, and the equality (\ref{equ_4}) follows from the requirement of security-level $r$, that is, the eavesdropper accessing any $r$ channels obtains no information about the message $M$.

Moreover, before discussing further, we need the following lemma (see \cite[Lemma 2]{Cai-Yeung-SNC-IT} and \cite{Han}).
\begin{lemma}\label{lem_Han}
For a subset $\alpha$ of $\mN=\{1,2,\cdots, n\}$, let $\bar{\alpha}=\mN\backslash \alpha$ and denote $(X_i:\ i\in \alpha)=X_{\alpha}$. For any $1\leq r \leq n$, let $$h_r=\frac{1}{{n-1\choose r-1}}\sum_{\alpha: |\alpha|=r}H(X_{\alpha}|X_{\bar{\alpha}}).$$
Then $h_1\leq h_2 \leq \cdots \leq h_n$.
\end{lemma}

Recall $H(M)=I(Y_{U\backslash J}; M|Y_J)$ for any $J\subset U$ with $|J|=r$. Summing over all $J\subset U$ with $|J|=r$, we deduce
\begin{align*}
&{\w+\bar{r}_K^* \choose r}H(M)\\
=&\sum_{J\subset U:\ |J|=r} I(Y_{U\backslash J}; M|Y_J)\\
=&\sum_{J\subset U:\ |J|=r} \big[H(Y_{U\backslash J}|Y_J)-H(Y_{U\backslash J}|M,Y_J)\big]\\
\leq& \sum_{J\subset U:\ |J|=r} H(Y_{U\backslash J}|Y_J)\\
=&{\w+\bar{r}_K^*-1 \choose \w+\bar{r}_K^*-r-1}
                              \cdot\left[ \frac{1}{{\w+\bar{r}_K^*-1 \choose \w+\bar{r}_K^*-r-1}}  \sum_{J\subset U:\ |J|=r} H(Y_{U\backslash J}|Y_J)\right]\\
\leq& {\w+\bar{r}_K^*-1 \choose \w+\bar{r}_K^*-r-1}H(Y_U),
\end{align*}
where the last inequality follows from Lemma \ref{lem_Han}. Hence,
\begin{align*}
H(Y_U) \geq \frac{{\w+\bar{r}_K^* \choose r}}{{\w+\bar{r}_K^*-1 \choose \w+\bar{r}_K^*-r-1}}H(M)= \frac{\w+\bar{r}_K^*}{\w+\bar{r}_K^*-r}H(M).
\end{align*}
Therefore, one has $H(M)+H(K)\geq H(M, K)$ and
\begin{align*}
H(M, K)=H(M,K,Y_U)\geq H(Y_U)\geq \frac{\w+\bar{r}_K^*}{\w+\bar{r}_K^*-r}H(M).
\end{align*}
Equivalently,
\begin{align}\label{equ_5}
H(K) \geq \frac{r}{\w+\bar{r}_K^*-r}H(M)=\frac{r\w}{\w+\bar{r}_K^*-r} \log q.
\end{align}
Since we have known that $r_K^*\leq r$, which further implies $\bar{r}_K^*\leq r$. Together with the inequality (\ref{equ_5}), we conclude that
$$H(K)\geq \frac{\w r}{\w+\bar{r}_K^*-r} \log q \geq r \log q,$$
which means that
$$r_K^*=\frac{H(K)}{\log q}\geq \frac{ r \log q}{\log q}=r.$$
Combining the above proof for two directions, we conclude $r_K^*=r$, which completes the proof.
\end{IEEEproof}
\begin{rem}
The above conclusion implies that even if for some reasons the source node gives up a part of network capacity or cannot send the maximum number of information symbols, then the number of random symbols won't be reduced. Actually, this can be interpreted intuitively as follows: if we have a safe that needs $3$ locks to be safe and secure, it does not matter how much plain document you are putting in the safe, the safe still needs $3$ locks and $3$ keys to be secured. The same understanding applies here.
\end{rem}

Actually, the proof of the achievability in Theorem \ref{thm_opt} induces a similar construction of SLNCs with security-level $r$ and no matter what the information rates are. Particularly, we can set $\vc=\vec{0}$, an all zero $(C_{\min}-\w-r)$-row vector for simplify. In addition, it also implies that the cardinality of the $\wtE_r$ is sufficient for constructing SLNCs. But we can see that $\wtE_r$ depends on the underlying LNC and it is hard to handle. Naturally, we hope to find another lower bound just depending on network topology, even though it is possibly looser. We define a new collection of channel-sets as:
$\wtE_r^{\cut} \triangleq \{ A\subseteq E:\ |A|=\mincut(s, A)=r \},$
where $\mincut(s, A)$ represents the minimum cut capacity between $s$ and $A$, clearly which just depends on the network topology. Further it is implied that $|\wtE_r^{\cut}|$ is a new lower bound.
\begin{prop}
$\wtE_r \subseteq \wtE_r^{\cut}$ and $|\wtE_r|\leq |\wtE_r^{\cut}| \leq { |E| \choose r}$.
\footnote{A better lower bound on field size for SLNCs has been obtained in our previous works. Please refer to \cite{Guang-SmlFieldSize-SNC-comm-lett} for more details.}
\end{prop}
\begin{IEEEproof}
For every $A\in \wtE_r$, we have $\Rank(F_A)=r$, which implies $\mincut(s, A)\geq \Rank(F_A)=r$. Together with $|A|=r$, one obtains $A\in \wtE_r^{\cut}$. The second inequality is obvious.
\end{IEEEproof}
\subsection{Imperfect Security Case}
The results in the last subsection can be extended to imperfect security case easily. First, we review the concept of imperfect security \cite[p.71]{Yeung-book}, which allows the eavesdropper to obtain a controlled amount of information about the source message. Again let $\mA$ be a collection of all channel-sets with cardinalities not larger than $r$. The perfect security condition $I(M;Y_A)=0$ for all $A\in \mA$ is replaced by the imperfect security condition $I(M;Y_A)\leq i\log q$ for all $A\in \mA$ (also see \cite{Cai-Yeung-SNC-IT}), where $i$ is a fixed nonnegative integer satisfying $0 \leq i \leq r$. This integer $i$ indicates that how much information can be leaked to the eavesdropper, and we call it as \textit{imperfect-security-level}. For imperfect security, we can still claim that the minimum number of random key won't decrease, no matter how much information can be transmitted to sinks. In the following, we state our result.

\begin{thm}
Let the wiretap collection $\mA$ consist of all channel-sets with cardinalities not larger than $r$ and the required imperfect-security-level be $i$. Whatever the information rates $\w$ satisfying $\w+r-i< C_{\min}$ or equivalently $\w< C_{\min}-r+i$ are, the optimal rate $r_K^*$ of the key is $r-i$.
\end{thm}

The proof of this theorem is closely similar to that of Theorem \ref{thm_opt}, thus omitted.

\section{Conclusion}

In this letter, we focus on the minimum number of random key injected into networks by the source node to prevent information from being leaked to an eavesdropper which can access any $r$ channels at most, no matter how much source message is multicast to all sinks. For two cases of perfect and imperfect securities, we respectively deduce two minimums and further show that they just depend on the security constraint and independent to other parameters.


%





\ifCLASSOPTIONcaptionsoff
  \newpage
\fi



%

%

\end{document}